\documentclass[twocolumn,prl,showpacs]{revtex4}

\usepackage{epsfig}

\begin{document}

\title{Entanglement of Two Impurities through Electron Scattering}

\author{A. T. Costa Jr.}
\affiliation{Departamento de Ci\^encias Exatas,
Universidade Federal de Lavras, 37200-000 Lavras, Brazil}
\author{S. Bose}
\affiliation{Department of Physics and Astronomy, University
College London, Gower Street, London WC1E 6BT, United Kingdom}
\author{Y. Omar}
\affiliation{Centro de F\'isica de Plasmas, Instituto Superior T\'ecnico,
P-1049-001 Lisbon, Portugal}

\date{22 March 2005}


\begin{abstract}

We study how two magnetic impurities embedded in a solid can be
entangled by an injected electron scattering between them and by
subsequent measurement of the electron's state. We start by
investigating an ideal case where only the electronic spin
interacts successively through the same unitary operation with the
spins of the two impurities. In this case, high (but not maximal)
entanglement can be generated with a significant success
probability. We then consider a more realistic description which
includes both the forward and back scattering amplitudes. In this
scenario, we obtain the entanglement between the impurities as a
function of the interaction strength of the electron-impurity
coupling. We find that our scheme allows us to entangle the
impurities maximally with a significant probability.
\end{abstract}


\maketitle


 Recently there has been an increasing interest on the generation
of entanglement among spins in mesoscopic solid state structures
\cite{stationary,kane,yamamoto,mobile,antonio,bose-home,saraga,beenakker1,beenakker}.
While there are several schemes for entangling {\em adjacent}
stationary spins through a direct quantum gate between them
\cite{stationary,kane,yamamoto}, there is an unfortunate dearth of
schemes for entangling well separated stationary spins in such
structures. An overwhelming majority of the proposed schemes in
which one obtains a reasonable separation between the entangled
spins are for {\em mobile} entities
\cite{mobile,antonio,bose-home,saraga,beenakker1,beenakker}.
Entangling well separated stationary spins is practically
important because they can be constituents (qubits) of distinct
quantum computers. Establishing entanglement between them is
equivalent to linking these computers. Even if they themselves are
not parts of quantum computers, they can each have a switchable
interaction with static spin qubits of well separated quantum
computers. In absence of a method of entangling well separated
stationary spins, one has to design a scheme to either stop mobile
spins after they have traversed a distance, or find a scheme for
mapping their state on to stationary spins. In addition to the
above pragmatic application, such entangled stationary spins will
also enable one to test Bell's inequalities with massive
particles, which is yet to be done for a significant separation
\cite{trap-bell}. Of course, mobile entangled electrons can also
enable such tests with the individual spins being measured by spin
filters or spin selective detectors \cite{kawabata,fazio}. There
are also proposals for Bell's inequality measurements using the
orbital or path, as opposed to the spin, degree of freedom of
mobile entities \cite{ionicou,beenakker,buttiker}. However, quite
a few of the proposals for the measurement of a single spin, such
as those based on scanning tunnelling microscopy or magnetic
resonant force microscopy are specific to stationary spins
\cite{single-spin-meas}. These spin measurement methods could be
used in a Bell inequality experiment if one were able to entangle
well separated stationary spins.

   With the above motivations is mind, in this article we propose a
scheme to entangle two magnetic impurities (stationary spins
$1/2$) embedded in a solid state system. The main idea is to use a
ballistic electron as an agent which scatters off the two
impurities in succession and entangles them. Being a scattering
based scheme, it requires no control over the ability to switch
interactions on and off between entities in a solid, as is
required by many existing entangling proposals \cite{mobile}.
Moreover, even in comparison to other reduced control proposals,
such as those based on scattering or two particle interference
\cite{antonio,bose-home}, our current scheme has the simplicity
that it involves only one mobile entity, namely the ballistic
electron, and does away with the difficulty of having to make two
electrons coincide at the same place at the same time.

\begin{figure}[ht]
\begin{center}
\epsfig{file=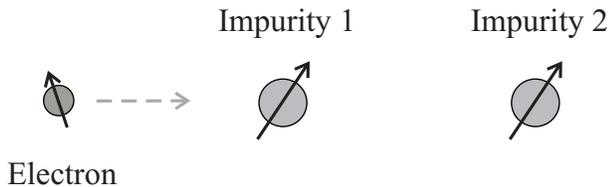, width=8cm}
\end{center}
\caption{Setup for our scheme to entangle two magnetic impurities
of spin $1/2$ in a solid through electron scattering. We consider
the simple case of a one-dimensional metallic atomic chain (of
non-magnetic atoms) where the two impurities are embedded. The
electron flies along the chain and its spin interacts with the
spins of the impurities, scattering the electron off and
entangling the two impurities.} \label{twoimpuritiesdiagram}
\end{figure}

   We comment first on the geometry of the system. Since
entanglement generation depends on a conduction electron
interacting with both impurities, it is most convenient to make
the system's cross section as small as possible. In this spirit,
and for the sake of simplicity, we consider a one-dimensional
metallic atomic chain (of non-magnetic atoms), with two embedded
(substitutional) spin-$1/2$ magnetic impurities. This is shown in
Fig.\ \ref{twoimpuritiesdiagram}.


 We know that in an ideal case,
where a mediating agent is allowed to interact with two systems
through distinct unitary operations, it can then perfectly
entangle them. The first of these unitaries perfectly entangles
the first system with the agent, and then the second operation
swaps the state of the agent with that of the second system. This
technique has, for example, been used in proposals for entangling
the state of cavities using flying atoms \cite{haroche}. The
different unitaries are implemented by different interaction times
or strengths between the agent and each of the systems. Such a
technique obviously requires either a great control over the
motion of the agent, or the non-trivial engineering of different
interaction strengths of the agent with the systems. Under these
circumstances, it becomes interesting to investigate the reduced
control situation where an agent interacts with both systems
through the same unitary operation. How well can the systems be
entangled under these circumstances?  In the context of quantum
optics, one can think of this as an atom having the same flight
time and interaction strength with two cavities through which it
flies in succession. An analog of this scenario in solid state
systems would be to have an electron flying past two identical
impurities with the same velocity, interacting magnetically with
them successively without any scattering. We first consider this
simplified case, just in order to investigate how much
entanglement can be established between two impurities, even when
the electron interacts with them through the same unitary. We find
that a significant entanglement can be established with a
significant probability as long as we can make an appropriate
measurement of the state of the electron after its successive
interaction with the two impurities. This case may not be
realistic from the solid state scattering scenario, but it is an
interesting precursor to the case when spatial scattering is
involved. Moreover, it can be realized in a quantum optics setting
where the electronic spin states are replaced by the atomic
internal states and impurity states are replaced by zero and one
photon states in the cavities. We then proceed to the realistic
case of the electron being spatially scattered by the interaction
with the impurities. Interestingly, in this case, we find that the
electron can entangle the two impurities near perfectly
(conditional on a favorable outcome of a measurement of the
electron's spin). Moreover, the probability of this favorable
outcome is significant (above 40 percent), which means that on
average one should be able to perfectly entangle the impurities
with three attempts.

\begin{figure}[ht]
\begin{center}
\epsfig{file=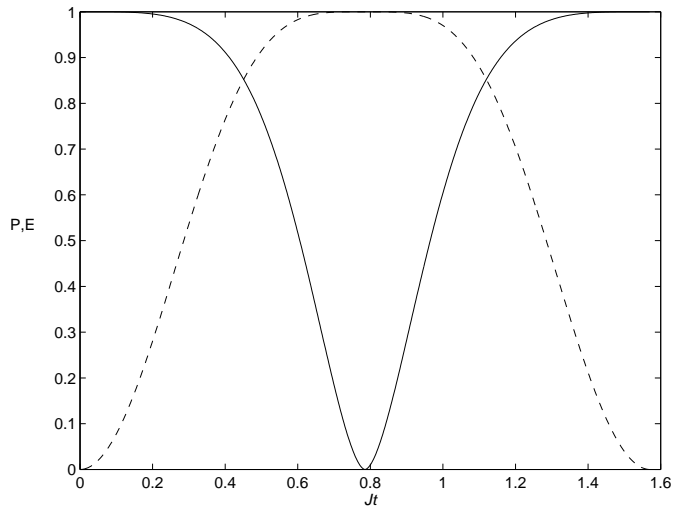, width=3.5in}
\end{center}
\caption{Ideal case --- Plots of the probability of success $P$
(dashed line) of detecting the transmitted electron in the spin
down state, and the entanglement $E$ (solid line) obtained between
the impurity spins in that case, as a function of $Jt$, for the
ideal scenario where the electron successively interacts with the
two impurities through the same unitary operation. It is clear
that one can obtain a high amount of entanglement with a
significant probability: for example, an entanglement of $0.99$
with a probability of $0.41$, or an entanglement of $0.84$ with a
probability of $0.86$.} \label{Fig. Ideal}
\end{figure}

    We begin by considering the ideal scenario where the
    electron's spin
    interacts in succession with each of the impurity spins through the
    Hamiltonian
    \begin{equation}
    H=J\overrightarrow{S}.\overrightarrow{\sigma},
    \end{equation}
    where $\overrightarrow{\sigma}$ refers to the Pauli
    operators of the electronic spin,
    $\overrightarrow{S}$ refers to Pauli
    operators for the impurity spins and $J$ is the coupling constant between the spins. We now assume that the
    electron interacts with the two impurities in succession for
    equal intervals of time, so that with both impurities the
    same unitary operation is implemented. The joint unitary operation between the electron
    and impurity $\iota$ (with $\iota=1,2$) as a function of the interaction time $t$ is
    given by (expressing the unitary operation in terms of its eigenstates):
    \begin{eqnarray}
    U_{e\iota}(t)&=&e^{-i t H}\nonumber\\
    &=& e^{i3Jt}|\Psi^{-}\rangle\langle\Psi^{-}|_{e\iota}\nonumber\\
    &+& e^{-iJt} \left(| \! \uparrow\uparrow\rangle\langle
    \uparrow\uparrow \! | + | \! \downarrow\downarrow\rangle\langle
    \downarrow\downarrow \! | \right)_{e\iota}\nonumber\\
    &+& e^{-iJt}|\Psi^{+}\rangle\langle\Psi^{+}|_{e\iota},
    \end{eqnarray}
where
\begin{equation}
|\Psi^{\pm}\rangle_{e\iota} = \frac{1}{\sqrt{2}} \left(| \!
\uparrow\downarrow\rangle_{e\iota} \pm| \!
\downarrow\uparrow\rangle_{e\iota}\right).
\end{equation}
Let us consider the following initial state where the impurity
spins are aligned, and the electron's spin is anti-aligned with
them
\begin{equation}
| \psi_0 \rangle = | \! \uparrow\rangle_{e}| \!
\downarrow\rangle_{1}| \! \downarrow\rangle_{2}.
\end{equation}
The final state is then given by
\begin{eqnarray}
&& | \psi_f \rangle = U_{e2}(t) U_{e1}(t) | \psi_0 \rangle = \\
&& \frac{e^{-i2Jt}}{2} \left( \alpha | \! \uparrow\rangle_{e}| \!
\downarrow\rangle_{1}| \! \downarrow\rangle_{2} + \beta | \!
\downarrow\rangle_{e}| \! \uparrow\rangle_{1}| \!
\downarrow\rangle_{2} + \gamma | \! \downarrow\rangle_{e}| \!
\downarrow\rangle_{1}| \! \uparrow\rangle_{2} \right), \nonumber
\end{eqnarray}
where $\alpha=(1+e^{i4Jt})^2/2$, $\beta=1-e^{i4Jt}$ and
$\gamma=(1-e^{i8Jt})/2$. Each interaction either leaves the
direction of the spins unchanged, either flips the interacting
pair. Note that if we now measure the spin of the electron and
observe the state $| \!\! \downarrow\rangle_{e}$, the impurities
will be left in the entangled state $\beta | \!\!
\uparrow\rangle_{1}| \!\! \downarrow\rangle_{2} + \gamma | \!\!
\downarrow\rangle_{1}| \!\! \uparrow\rangle_{2}$. In Fig.\
\ref{Fig. Ideal} we present the probability of this outcome
(dashed line), as well as the resulting amount of entanglement
quantified by the entanglement of formation \cite{entform} (solid
line) between the two impurities, both as a function of $Jt$, the
product of the interaction strength and the interaction time. We
study the probability and the entanglement as a function of $Jt$
in the interval $[0, \pi/2]$ as they are periodic functions, and
observe that in this ideal model maximal entanglement is generated
only with zero probability. This, however, does not rule out the
possibility of obtaining a high amount of entanglement with a
significant probability: for example, an entanglement of $0.99$
with a probability of $0.41$, or an entanglement of $0.84$ with a
probability of $0.86$ as seen from Fig. \ref{Fig. Ideal}.


   Let us now move to a more realistic scattering scenario.
Magnetic impurities embedded in a conduction electron sea are
traditionally modelled by a s-d Hamiltonian \cite{hewson}. In this
model the magnetic impurities are localized spins interacting with
the conduction electrons via an exchange term. The full
hamiltonian of a system with one impurity reads
\begin{equation}
H = \sum_{k,\sigma}\varepsilon_{k}a^\dagger_{k\sigma}a_{k\sigma} +
\sum_{kk'}J_{kk'}\vec{S}.\vec{s}_{kk'}\, ,
\end{equation}
where $\vec{S}$ is the impurity spin operator, $a^\dagger_{k\sigma}$
creates an electron with wavevector $k$ and spin $\sigma$ and
\begin{equation}
\vec{s}_{kk'} = \hat{a}^\dagger\vec{\sigma}\hat{a}\, ,
\end{equation}
with $ \hat{a} = \left( \begin{array}{c}
                         a_{k\uparrow} \\
                         a_{k\downarrow}
                 \end{array} \right).
$
The s-d Hamiltonian is actually derived from the more fundamental Anderson
Hamiltonian through the Schrieffer-Wolff transformation. As a consequence,
the interaction strength $J$ is related to the strength of the Coulomb
interaction between electrons and the hybridization of narrow and conduction
bands \cite{hewson}. In our calculation we will adopt the usual assumption
that $J$ is independent of $k,k'$.

We want to find out how much entanglement may be generated by a conduction
electron that is injected in the system and interacts with both magnetic
impurities. One may determine the system's final state by calculating the
scattering matrix associated with each impurity and combining them together.
The result is a sequence of (infinitely many) scattering processes, in which
the output of a scattering event is the input of the subsequent one.
The result of each individual scattering process is determined by use of
Fermi's golden rule. The relevant $T$-matrix is calculated to first order
in the interaction.

If we consider that the conduction electron is being injected under
low bias, its energy and wavevector will be the Fermi energy and Fermi
wavevector of the system, respectively. We thus assume a initial state
of the form
\begin{equation}
|\psi_{in}\rangle = |k_F,\uparrow\rangle_e
|\!\downarrow\downarrow\rangle_{12}\, ,
\end{equation}
that represents a conduction electron with positive Fermi wavevector $k_F$
and spin $\uparrow$ along the quantization axis ($z$, for instance)
propagating towards the two impurities, whose spins are both $\downarrow$
along the $z$-axis.

As a result of the multiple scatterings of the conduction electron by the two
impurities, a final state is generated which is a superposition of
states in which the conduction electron has been reflected ($r$) or
transmitted ($t$),
\begin{equation}
|\psi_{out}\rangle = |\psi_{out}^r\rangle + |\psi_{out}^t\rangle\,
,
\end{equation}
and the transmitted component reads
\begin{eqnarray}
&& |\psi_{out}^t\rangle= \\ && A|k_F,\uparrow\rangle_e
|\!\downarrow\downarrow\rangle_{12} + B|k_F,\downarrow\rangle_e
|\!\uparrow\downarrow\rangle_{12} + C|k_F,\downarrow\rangle_e
|\!\downarrow\uparrow\rangle_{12}\, . \nonumber
\end{eqnarray}
The coefficients $A$, $B$ and $C$ may be expressed as an infinite
sum of powers of the product $J\rho(\varepsilon_F)$, which,
according to our estimates \cite{antonio}, is of the order of
unit. We verified numerically that the series converges rapidly
for $J\rho(\varepsilon_F)\in [0,2]$. Below we present the series
up to sixth order in $J\rho(\varepsilon_F)$ (corresponding to
three iterations of the scattering matrix),
\begin{eqnarray}
A^{(3)} = \frac{1}{N}(t^2 + t^2\lambda^2-8t\lambda^3 + 16\lambda^6-7t^2\lambda^4)\nonumber\\
B^{(3)} = \frac{1}{N}(-2\lambda t + 2t\lambda^3-2t^2\lambda^2+ 6t\lambda^5+8t^2\lambda^4 )\nonumber\\
C^{(3)} = \frac{1}{N}(-2\lambda t + 8\lambda^4-2t\lambda^3+2t^2\lambda^2 + 6t\lambda^5)\, ,
\end{eqnarray}
where $\lambda = \pi iJ\rho(\varepsilon_F)/2$, $t=1-\lambda$ and $\frac{1}{N} = \sqrt{|A^{(3)}|^2 + |B^{(3)}|^2 + |C^{(3)}|^2}$.

\begin{figure}[ht]
\begin{center}
\epsfig{file=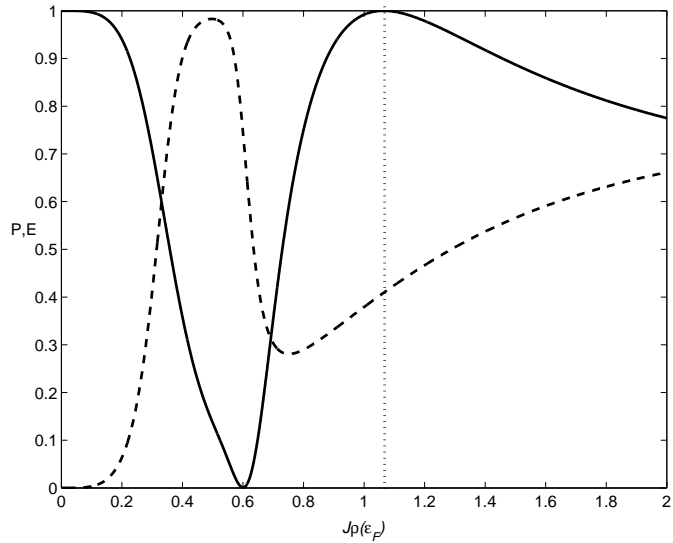, width=3.5in}
\end{center}
\caption{Realistic case --- Plots of the probability $P$ (dashed
line) of detecting the transmitted electron in the spin down
state, and the amount of entanglement $E$ (solid line) obtained
between the impurity spins in that case,  as a function of the
interaction strength $J\rho(\epsilon_F)$, for the realistic
scenario of the electron successively scattering off the two
impurities. It is clear that at certain values of
$J\rho(\epsilon_F)$, as shown by the dotted vertical line, the
impurities can be projected on to a maximally entangled state with
a significant ($0.41$) probability of success.} \label{Fig. Real}
\end{figure}

We now proceed to calculate the amount of entanglement (as
quantified by the entanglement of formation) generated
\textsl{conditional on an electron being transmitted}, which is
the entanglement contained in the state $|\psi_{out}^t\rangle$.
Notice that if the transmitted electron has spin up the final
state has zero entanglement. Thus we will only evaluate the
entanglement of the state in which the transmitted electron has
spin down. Fig.\ \ref{Fig. Real} shows the entanglement in this
state (solid line) and the probability $P$ of observing a
transmitted electron with spin up (dashed line). One may notice
that there is some entanglement for most of the range
$0<J\rho(\varepsilon_F)<2$, and the probability $P$ is also
considerable. Moreover, there are values of $J\rho(\varepsilon_F)$
for which the entanglement is maximum, and $P$ is significant
($0.41$). It is somewhat interesting to note that although in the
ideal case one cannot prepare the impurities in a maximally
entangled state with a non-vanishing probability, one can do so in
the realistic case.


    In this article, we have presented a scheme for entangling two
magnetic impurities in a solid through the scattering of a single
ballistic electron. While much work has been done on entangling
spins in mesoscopic solid state systems, this is the first
proposal for entangling distant stationary spins without the aid
of an array of intervening spins. An immediate consequence will be
in testing Bell's inequalities with well separated stationary
spins in a solid (of course, we still have to measure the spin of
a mobile electron, but not when testing Bell's inequalities). A
more far reaching and more significant consequence will be in
interfacing distant spin quantum computers. Our work shows that
maximal entanglement can be obtained between the distant
stationary spins with a significant probability. The scheme should
be implementable using the same systems as those used to study
Kondo physics \cite{kondo}. In the future, it will be interesting
to explore the types of multi-particle entangled states that can
be obtained by the scattering of a single electron from a series
of such impurities.

\begin{acknowledgments}

SB and YO wish to thank the Institute for Quantum Information at
Caltech, where this work was started, for their hospitality. ATC
acknowledges financial support from CNPq (Brazil). SB acknowledges
the EPSRC QIPIRC. YO acknowledges financial support from
Funda\c{c}\~{a}o para a Ci\^{e}ncia e a Tecnologia (Portugal) and
the 3rd Community Support Framework of the European Social Fund,
and from FCT and EU FEDER through project POCTI/MAT/55796/2004
QuantLog.

\end{acknowledgments}


\end{document}